\documentclass[journal]{IEEEtran}

\ifCLASSINFOpdf
\else
   \usepackage[dvips]{graphicx}
\fi
\usepackage{url}

\usepackage{graphicx}

\begin{document}

\title{Depth extraction from a single compressive hologram}

\author{Baturay Ozgurun and Mujdat Cetin, \IEEEmembership{Fellow, IEEE}
\thanks{This work was not supported by any organization.}
\thanks{B. Ozgurun is with the Department of Biomedical Engineering, University of Rochester, Rochester, NY 14627, USA (e-mail: bozgurun@ur.rochester.edu).}
\thanks{M. Cetin is with the Department of Electrical and Computer Engineering, University of Rochester, Rochester, NY 14627, USA (e-mail: mujdat.cetin@rochester.edu).}
\thanks{B. Ozgurun is also with the Faculty of Engineering and Natural Sciences, Sabanci University, Istanbul 34956, Turkey}
\thanks{B. Ozgurun is also with the School of Engineering and Natural Sciences, Istanbul Medipol University, Istanbul 34810, Turkey}}

\maketitle

\begin{abstract}
We propose a novel method that records a single compressive hologram in a short time and extracts the depth of a scene from that hologram using a stereo disparity technique. The method is verified with numerical simulations, but there is no restriction on adapting this into an optical experiment. In the simulations, a computer-generated hologram is first sampled with random binary patterns, and measurements are utilized in a recovery algorithm to form a compressive hologram. The compressive hologram is then divided into two parts (two apertures), and these parts are separately reconstructed to form a stereo image pair. The pair is eventually utilized in stereo disparity method for depth map extraction. The depth maps of the compressive holograms with the  sampling rates of 2, 25,  and 50 percent are compared with the depth map extracted from the original hologram, on which compressed sensing is not applied. It is demonstrated that the depth profiles obtained from the compressive holograms are in very good agreement with the depth profile obtained from the original hologram despite the data reduction.
\end{abstract}

\begin{IEEEkeywords}
Stereo image processing, holography, compressed sensing.
\end{IEEEkeywords}

\IEEEpeerreviewmaketitle

\section{Introduction}

Holography is a technique to record and reconstruct a three-dimensional (3D) object. To record a hologram, a coherent light source is usually divided into two arms: reference and object beams. The object beam first illuminates the 3D object, and then it reflects toward a beam splitter. The beam carries phase and amplitude information related to the object. To extract the phase information, the reference beam is also required. Therefore, the reference and object beams are merged on the beam splitter. Interference of two beams, which is also called a hologram, is recorded with a camera for a numerical reconstruction.

To extract depth from the hologram, several methods have been developed. The Fresnel propagation method is one of the widely used techniques for hologram reconstruction. This method enables us to calculate the depth of microscopic objects; however, it requires numerical phase unwrapping when the phase is wrapped for distances longer than a wavelength \cite{ulf2005}. Phase unwrapping is useful for microscopic objects, but this is not sufficient for macroscopic objects because of their size. Phase shifting is another method for depth extraction, but limited depth of field restricts the depth acquisition for macroscopic objects \cite{yamaguchi2003,yamaguchi1997}. Other researchers have shown that a dual beam illumination can be utilized to obtain two phase-contrast images, and subtraction of these images can provide depth of macroscopic objects. However, 2$\pi$ jumps reduce the efficiency of this method \cite{prieto2006,solis2012}. In addition, it was demonstrated that the gray level variance method can be used to extract depth of macroscopic objects, but this technique works mostly when a highly textured object is used \cite{frey2007,ma2004,mcelhinney2007}.

Pitk\"{a}aho and Naughton presented a study for the depth extraction from a single hologram \cite{pitkaaho2011}. They sharply divided the single hologram along the horizontal direction into two separated holograms. Each separated hologram is equally sized with the single hologram but contains half of the intensity values of it. After intensity division, the separated holograms were independently reconstructed to form a stereo image pair. Eventually, the image pair was utilized in a stereo disparity method to get the depth information related to the object. We were inspired by the study of Pitk\"{a}aho and Naughton and we have recently demonstrated depth extraction for macroscopic objects from experimentally recorded holograms \cite{ozgurun2017}. We also have shown that depth extraction is mostly independent from the division directions (horizontal, vertical, and diagonal) as well as the division types (gradual and sharp). Although the study of Pitk\"{a}aho and Naughton as well as our previous study demonstrated that depth of small and macroscopic objects could be extracted from a single hologram, high-speed recording and high-speed depth extraction are still challenging problems because of huge data volumes \cite{blinder2019}. High-speed depth extraction could be possible with our previously proposed approach in \cite{ozgurun2017} because depth can be extracted from a single hologram alone, but there is also the desire to record a hologram in a short time. In this Letter, we propose a method that records a hologram faster using the compressed sensing (CS) framework and extracts depth information from a single compressive/estimated hologram using a stereo disparity method.

\section{Imaging via Compressed Sensing}

In conventional imaging, a camera records an image by sampling a scene at the Nyquist rate, and it collects $N$ measurements, where $N$  is the total number of image pixels. Once the image is recorded, it is usually compressed to reduce its dimension. To perform compression, the image is first represented as a sparse image by utilizing a sparsifying transform. Then, the most significant coefficients of the transform domain representation of the image are kept, and the rest of the coefficients are thrown away. Eventually, the approximated transform coefficients is back transformed while keeping only the most significant coefficients \cite{mallat2009}. Given that many coefficients are thrown away, one could argue sampling the scene at the Nyquist rate wastes hardware. The compressed sensing (CS) framework is quite different from conventional imaging. It does not have to meet the Nyquist sampling rate and then carry out compression operations, rather it needs only $M$ measurements, where $M \ll N$, to recover a scene. The CS framework has been utilized for a variety of applications such as radar imaging \cite{potter2010} and magnetic resonance imaging \cite{lustig2008}, but here we focus on the application of CS in optics. Once of the first applications of CS in optics was the development of a single-pixel camera \cite{duarte2008}. In that application, a camera is not used for recording a scene. The scene is first sampled with pseudorandom patterns, which are generated by a digital micromirror device (DMD). Generated random patterns construct a sampling matrix $\Phi$, where $\Phi \in \Re^{M \times N}$. The inner products between the random patterns and the scene are collected by a photodiode, which generates measurements $y$, where $y \in \Re^{M}$. If the scene is sparse enough, a sensing matrix $A$, where $A \in \Re^{M \times N}$, can be constructed only from the sampling matrix $\Phi$. However, if the scene is dense, the sensing matrix $A$ must be formed with the product of the sampling matrix and a sparsifying matrix $\Psi$, where $\Psi \in \Re^{N \times N}$. This operation can be described as $A=\Phi \Psi$. The sparsifying matrix $\Psi$ represents a scene sparsely in an appropriate transform domain. Once the measurements $y$ and the sensing matrix $A$ are formed, they are utilized in a non-linear recovery algorithm to estimate the scene $\hat{x}$, where $\hat{x} \in \Re^{N}$. Most recovery algorithms are based on an $\ell_{1}$ minimization problem. This can be mathematically described below with an assumption that measurements are corrupted by a bounded noise $n$, i.e. $y=Ax + n$, where $x \in \Re^{N}$ is the scene, and $\Vert n \Vert_{2} \leq \varepsilon$.

\begin{equation}
\hat{x}=\min_{x} \Vert x \Vert_{1} \; s.t. \; \Vert y - Ax \Vert_{2} \leq \varepsilon
\end{equation}

There are some restrictions for adapting of CS into an optical configuration. First, the scene image must be sparse or it must be represented as a sparse image. A dense image can be sparsified by utilizing sparsifying matrices. However, it should be considered that the level of sparsity of the image affects the performance of the recovery algorithm. Second, the sampling patterns must satisfy the restricted isometry property (RIP). Fortunately, the RIP constraint can be satisfied when the sampling patterns are made from the random binary patterns that can be easily generated by the DMD \cite{candes2008}.

In the literature, CS is usually applied to holography for image reconstruction and data security applications \cite{leportier2017,marim2010,di2012}. However, CS can also potentially enable one to record a scene or a hologram in a short time since it samples the scene with a DMD instead of a camera, and it requires a small number of measurements to recover the scene. The frame rate of a typical DMD on the market is almost 330 times higher than that of a camera. In addition, it was demonstrated that it may be possible to recover a scene with a sampling rate of only 2 percent \cite{duarte2008}. Considering of the frame rate of the DMD and the ability of CS for recovering the scene with few measurements, an optical configuration based on CS can record a scene 2 or 3 times faster. Here, we claim that data collection time of a single hologram can be reduced by utilizing CS. In addition to recording a single hologram faster, depth extraction is performed from the recorded single hologram using a stereo disparity method.

\section{Method}

\begin{figure}[htbp]
	\begin{minipage}[t]{1\linewidth}
 	  \centering
 	  \centerline{\includegraphics[width=7.9cm]{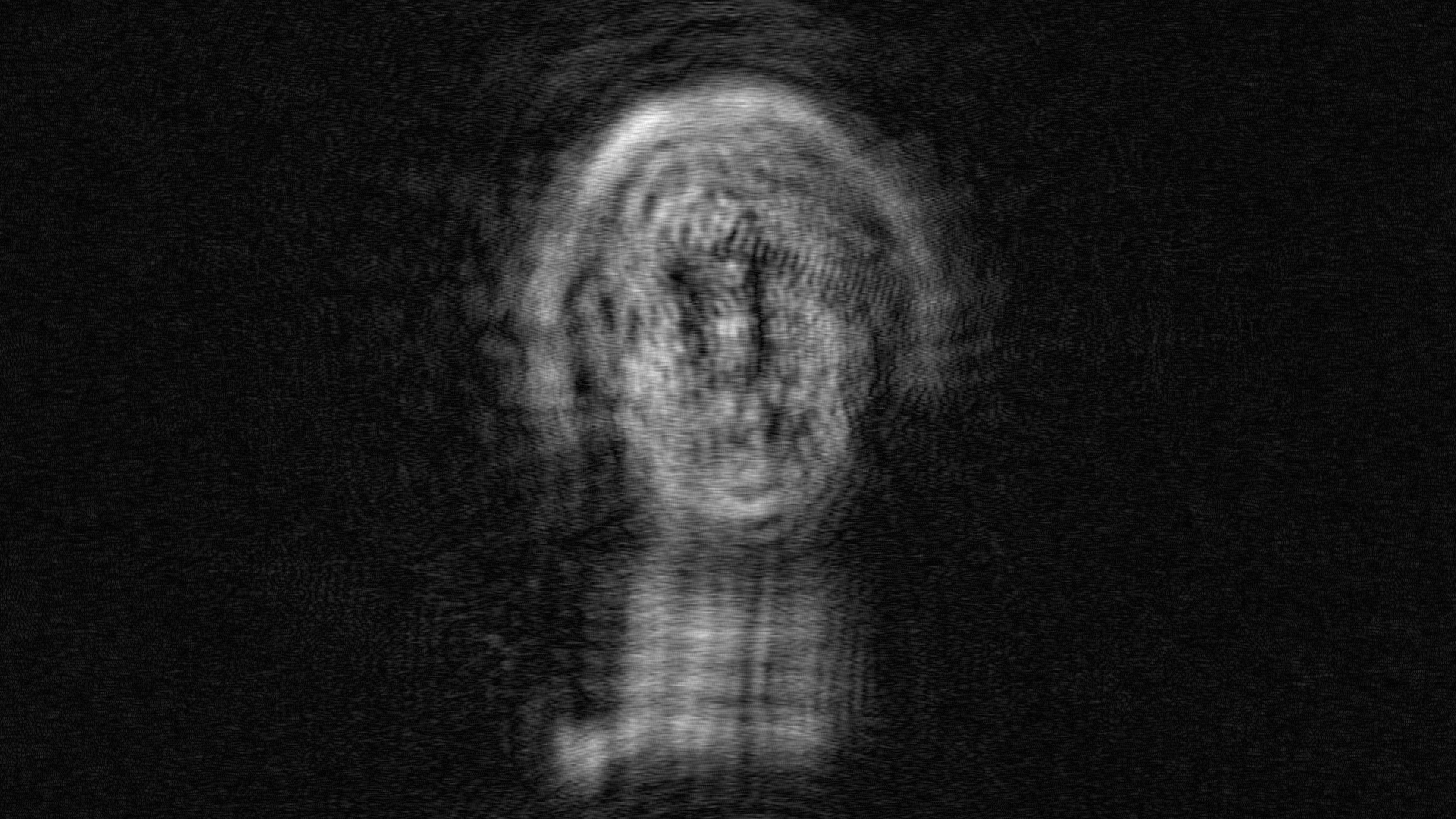}}
 	  \centerline{(a)}
	\end{minipage}
	\begin{minipage}[t]{1\linewidth}
 	  \centering
 	  \centerline{\includegraphics[width=7.9cm]{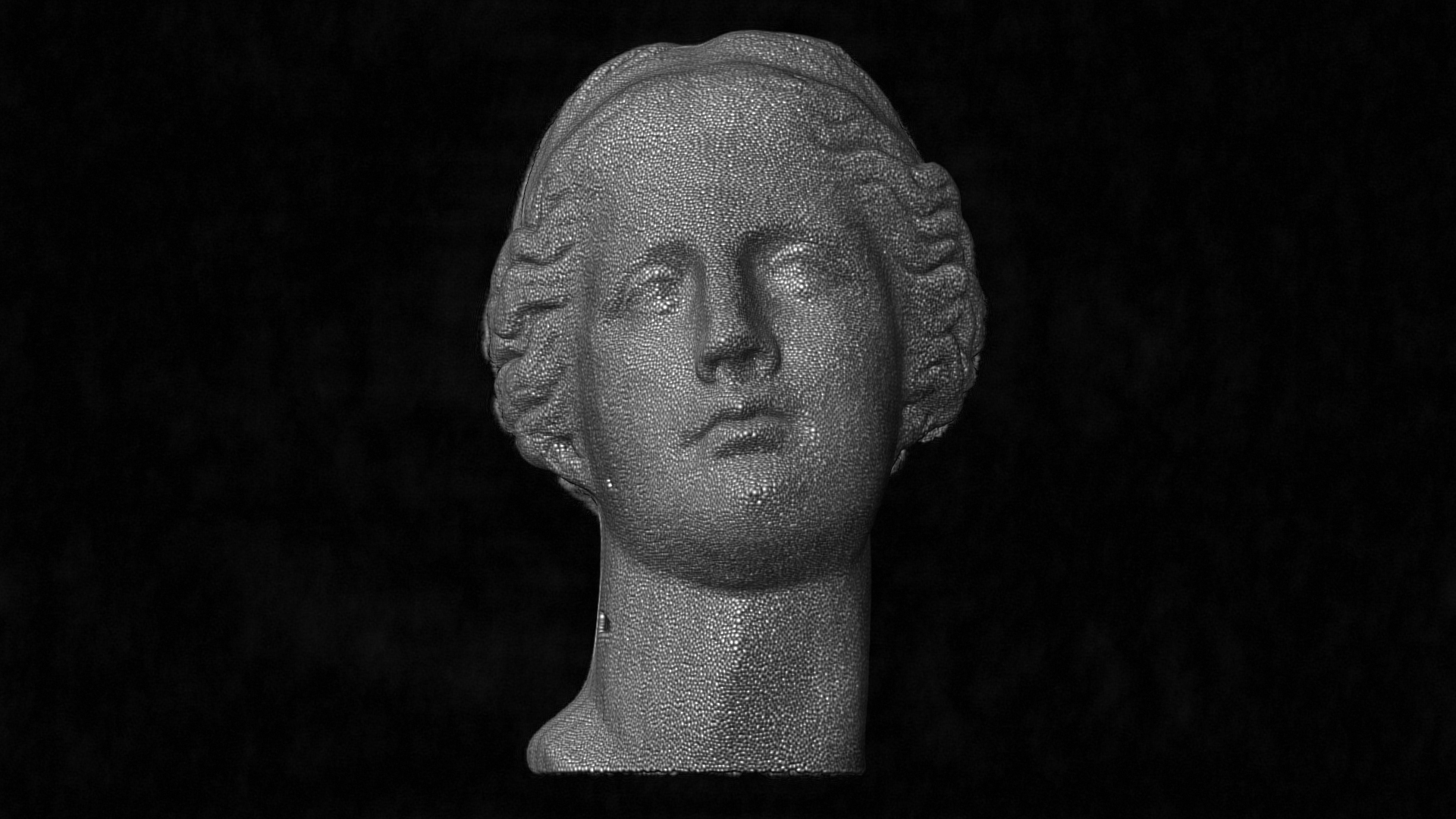}}
  	  \centerline{(b)}
	\end{minipage}
\caption{The CGH (a) of the Venus statue, and its numerical reconstruction (b) with the Fresnel approximation method.}
\label{fig:CGH}
\end{figure}

To demonstrate our claim, a computer-generated hologram (CGH) of the Venus statue, which is provided by David Blinder et al. as an open access file, is utilized \cite{blinder2015}. The CGH (1920 x 1080 pixels with a pixel pitch of 8 $\mu m$) and its numerical reconstruction with the Fresnel approximation method are presented in Fig. \ref{fig:CGH}. The data dimension of the CGH is high, and this increases the execution time. To reduce the computational cost, the CGH is first transformed into the Fourier domain, and the low frequency region (one-tenth of the bandwidth) is extracted and then back transformed. This operation compresses the hologram size by 100 times. Although the sharp transitions of the original hologram (1920 x 1080 pixels) disappear in the small hologram (192 x 108 pixels), most of the information about the structure of the Venus statue is preserved. In this study, this compressed hologram is used for all numerical calculations instead of the original hologram, and the small hologram is hereinafter referred to as the CGH.

\begin{figure}[htbp]
	\begin{minipage}[t]{0.495\linewidth}
 	  \centering
 	  \centerline{\includegraphics[width=4.25cm]{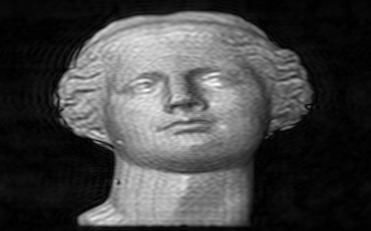}}
 	  \centerline{(a)}
	\end{minipage}
	\begin{minipage}[t]{0.495\linewidth}
 	  \centering
 	  \centerline{\includegraphics[width=4.25cm]{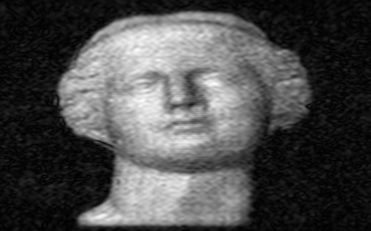}}
  	  \centerline{(b)}
	\end{minipage}
	\begin{minipage}[t]{0.495\linewidth}
 	  \centering
 	  \centerline{\includegraphics[width=4.25cm]{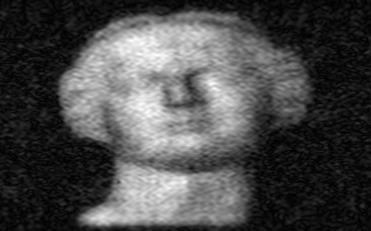}}
  	  \centerline{(c)}
	\end{minipage}
	\begin{minipage}[t]{0.495\linewidth}
 	  \centering
 	  \centerline{\includegraphics[width=4.25cm]{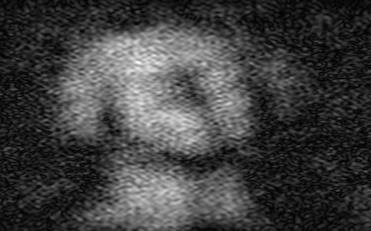}} 
  	  \centerline{(d)}
	\end{minipage}
\caption{The numerical reconstructions of the CGH (a) and the compressive holograms with the sampling rate of 50 percent (b), and 25 percent (c) and 2 percent (d). The reconstructions are performed with the Fresnel approximation method.}
\label{fig:CGHCS}
\end{figure}

In the simulation-based experiments, the CGH is considered as a holographic scene, and the DMD is considered to be placed in front of the CGH or a beam splitter that combines object and reference beams. The CGH is sampled with random binary patterns since the DMD can produce this type of patterns. Inner products between the random binary patterns and the CGH present measurements. In a real optical configuration, the measurements are usually collected by a photodiode or photomultiplier tube (PMT). A sensing matrix is constructed from the product of a sampling matrix and a sparsifying matrix. The sampling matrix is created from the random binary patterns while the discrete cosine transform (DCT) is selected as the sparsifying matrix. The measurements and the sensing matrix are utilized in the NESTA algorithm, which is one of the open source CS recovery algorithms \cite{becker2011}. The NESTA algorithm produces an estimated CGH or a compressive hologram. The CGH reconstruction and the reconstructions of the compressive holograms with sampling rates of 2, 25, and 50 percent are shown in Fig. \ref{fig:CGHCS}. All numerical reconstructions are performed with the Fresnel approximation method. The reconstruction result of the CGH is slightly better than the reconstruction results of the compressive holograms. We demonstrated that it is possible to record a hologram about 1.6 times faster. This corresponds to the 2 percent sampling rate case and assumes the frame rate of the DMD is 330 times higher than that of the camera.

\begin{figure}[hbt!]
	\begin{minipage}[t]{0.495\linewidth}
 	  \centering
 	  \centerline{\includegraphics[width=4.25cm]{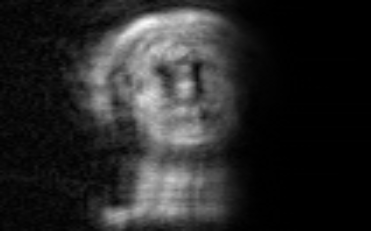}}
	\end{minipage}
	\begin{minipage}[t]{0.495\linewidth}
 	  \centering
 	  \centerline{\includegraphics[width=4.25cm]{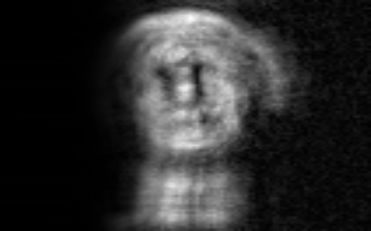}}
  	  \vspace{0.20cm}
	\end{minipage}
	\begin{minipage}[t]{0.495\linewidth}
 	  \centering
 	  \centerline{\includegraphics[width=4.25cm]{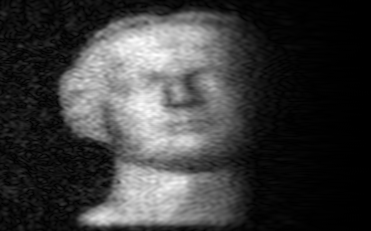}}
	\end{minipage}
	\begin{minipage}[t]{0.495\linewidth}
 	  \centering
 	  \centerline{\includegraphics[width=4.25cm]{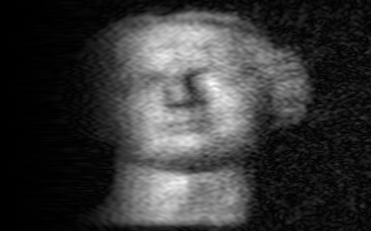}}
	\end{minipage}
\caption{The gradual intensity divisions of the compressive hologram (the first row) with the sampling rate of 25 percent, and their numerical reconstructions (the stereo image pair) with the Fresnel approximation method (the second row).}
\label{fig:HoloDiv}
\end{figure}

Once the compressive holograms are acquired, depth profiles are also obtained. We applied our previous study \cite{ozgurun2017}, which is based on the depth extraction from a single hologram, to the compressive holograms. To extract depth from a single compressive hologram, the compressive hologram is first divided gradually into two parts (two apertures) along the horizontal direction. Each of the separated holograms is equally sized with the single compressive hologram, but each of them contains almost half of the intensity weights of the single hologram. Division direction does not influence the accuracy of the depth information significantly; however, gradual division provides uniform illumination on the reconstruction, which increases the accuracy of the depth \cite{ozgurun2017}. After the hologram division is performed, two apertures are separately reconstructed with the Fresnel approximation method to form a stereo image pair. The stereo image pair and the separated holograms are presented in Fig. \ref{fig:HoloDiv}.

To extract depth, stereo disparity estimation is performed on the stereo image pair. The disparity technique generates depth map values, which are associated with depth of scene points and are usually shown as a gray-scale image. A small depth map value represents as a dark pixel in the gray-scale image and corresponds to a distant scene point. Similarly, a high depth map value or a bright pixel corresponds to a close scene point. In the literature, there exist various stereo disparity techniques. Here, we utilized the normalized cross-correlation (NCC) algorithm for the depth extraction, since this algorithm is robust to intensity offsets and contrast changes although it is computationally costly \cite{satoh2011}. The NCC algorithm calculates a correlation peak over two rectangular $(k \times k)$ blocks on the stereo image pair. These blocks are separately located on each stereo image pair, and they are called reference $R(x,y)$ and candidate $C(x,y)$ blocks. Calculation of the NCC is performed according to

\begin{equation}
NCC=\frac{\sum\limits_{x=1}^{k} \sum\limits_{y=1}^{k} \widetilde{R}(x,y) \widetilde{C}(x+\Delta,y)}{\sqrt{\sum\limits_{x=1}^{k} \sum\limits_{y=1}^{k} \widetilde{R}(x,y)^{2} \sum\limits_{x=1}^{k} \sum\limits_{y=1}^{k} \widetilde{C}(x+\Delta,y)^{2} }}
\end{equation}
where $\widetilde{R}(x,y) = R(x,y) - \overline{R}(x,y)$, $\widetilde{C}(x+\Delta,y) = C(x+\Delta,y) - \overline{C}(x,y)$, and $\Delta$ denotes any shifts applied. $\overline{R}(x,y)$ and $\overline{C}(x,y)$ are the mean pixel values over the reference and candidate blocks, respectively. Once the first NCC value is calculated $(\Delta=0)$, the candidate block is shifted one column for the second NCC calculation $(\Delta=1)$. The shifting operation is usually finalized when the shifting amount reaches half of the image size. This process provides a number of NCC values. The maximum value is picked and registered for the center pixel of the reference block. The overall operation must be repeated for the other pixels of the stereo image pair. This provides a depth map for the stereo image pair.

\section{Experimental Results}

\begin{figure}[htbp]
	\begin{minipage}[t]{0.495\linewidth}
 	  \centering
 	  \centerline{\includegraphics[width=5cm]{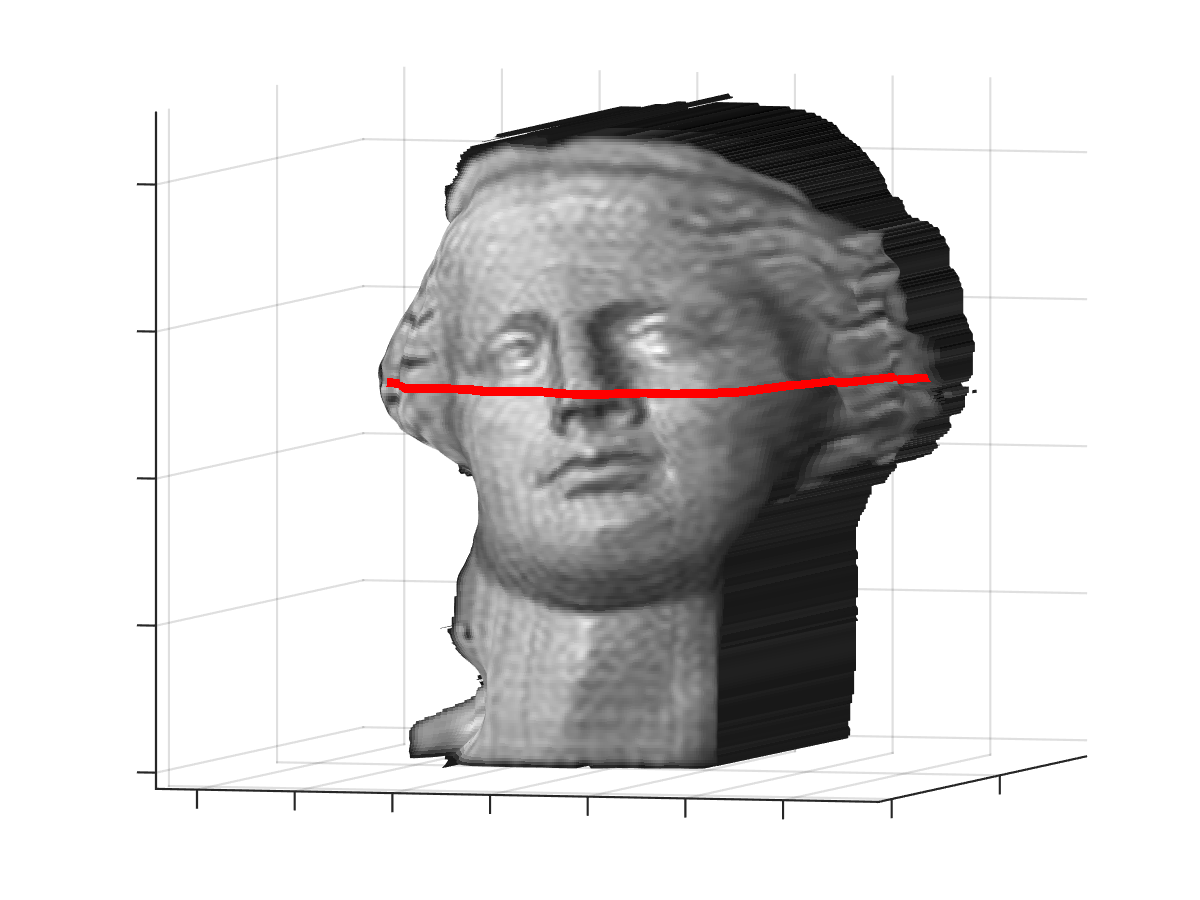}}
 	  \centerline{(a)}
	\end{minipage}
	\begin{minipage}[t]{0.495\linewidth}
 	  \centering
 	  \centerline{\includegraphics[width=5cm]{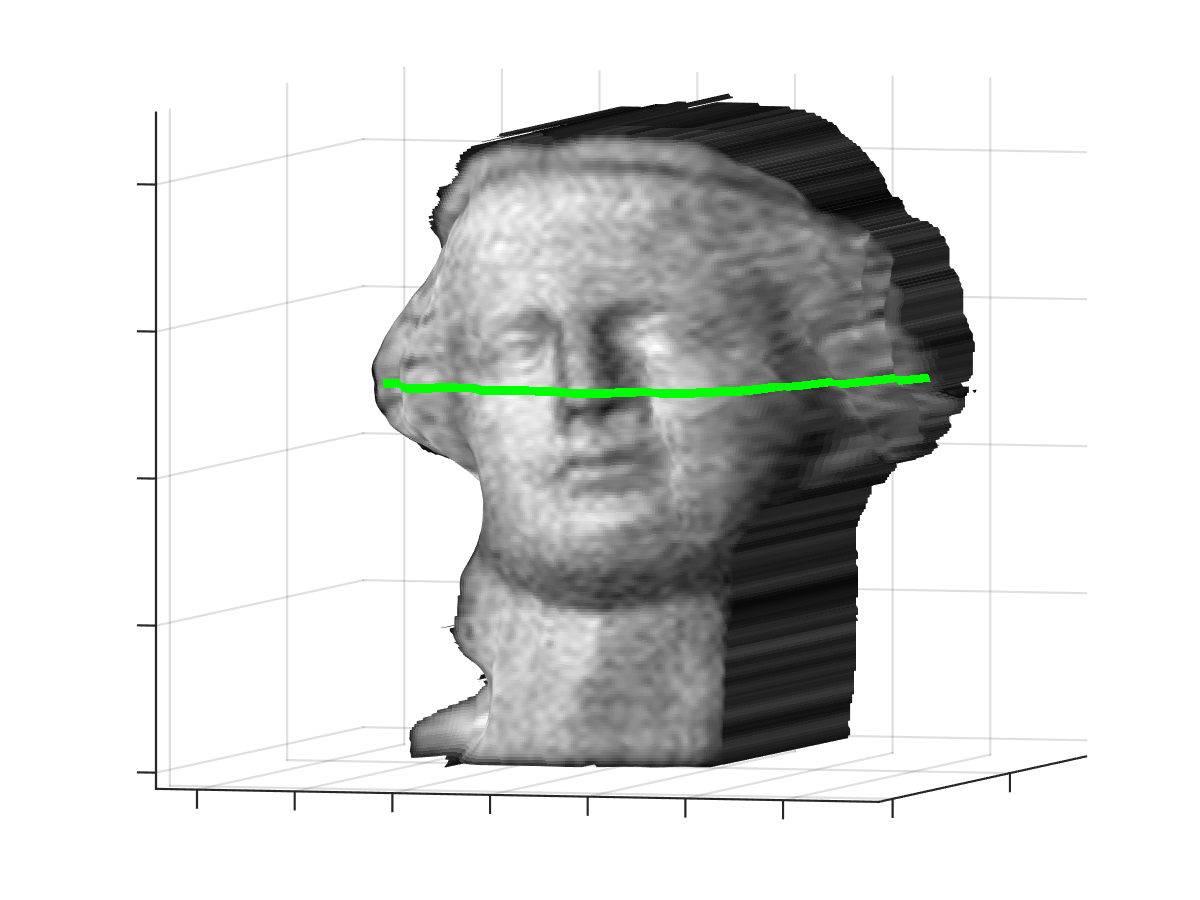}}
  	  \centerline{(b)}
	\end{minipage}
	\begin{minipage}[t]{0.495\linewidth}
 	  \centering
 	  \centerline{\includegraphics[width=5cm]{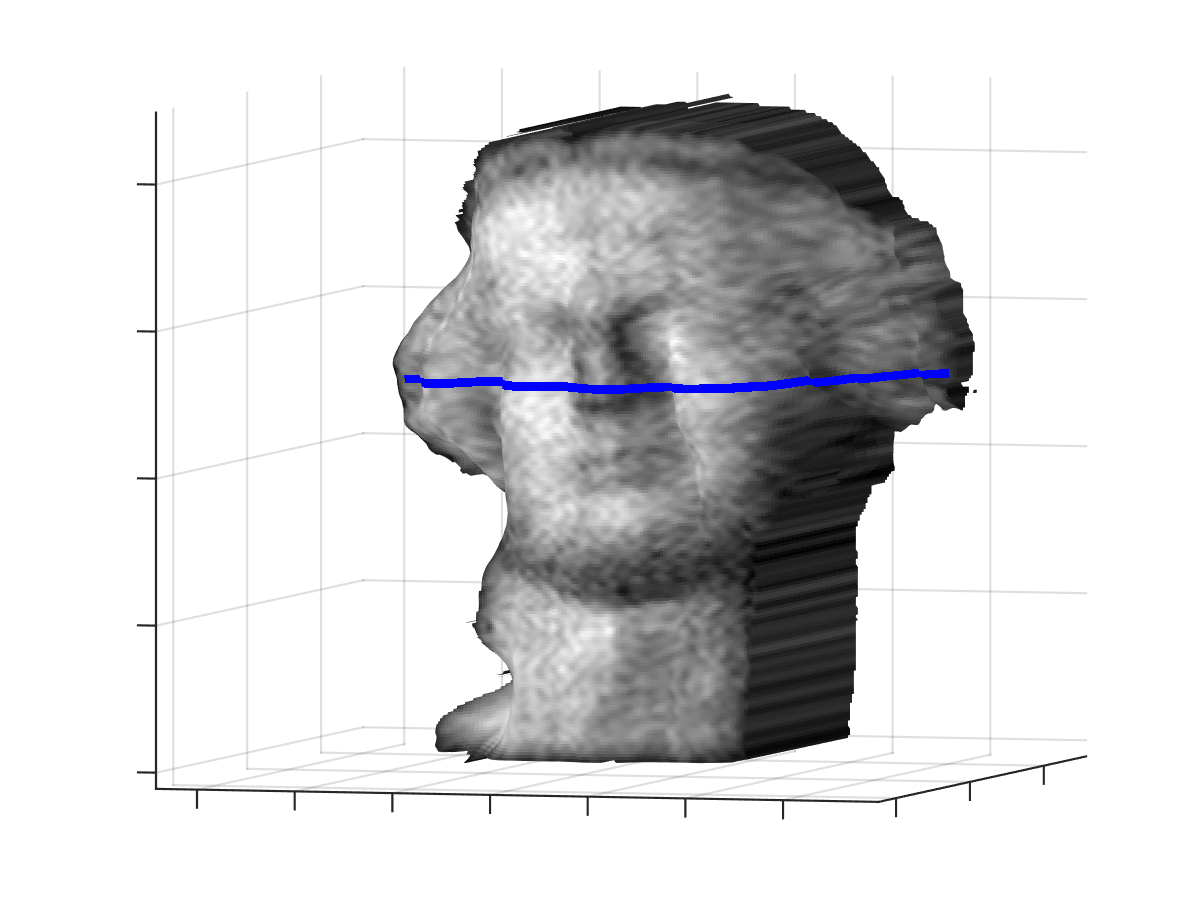}}
  	  \centerline{(c)}
	\end{minipage}
	\begin{minipage}[t]{0.495\linewidth}
 	  \centering
 	  \centerline{\includegraphics[width=5cm]{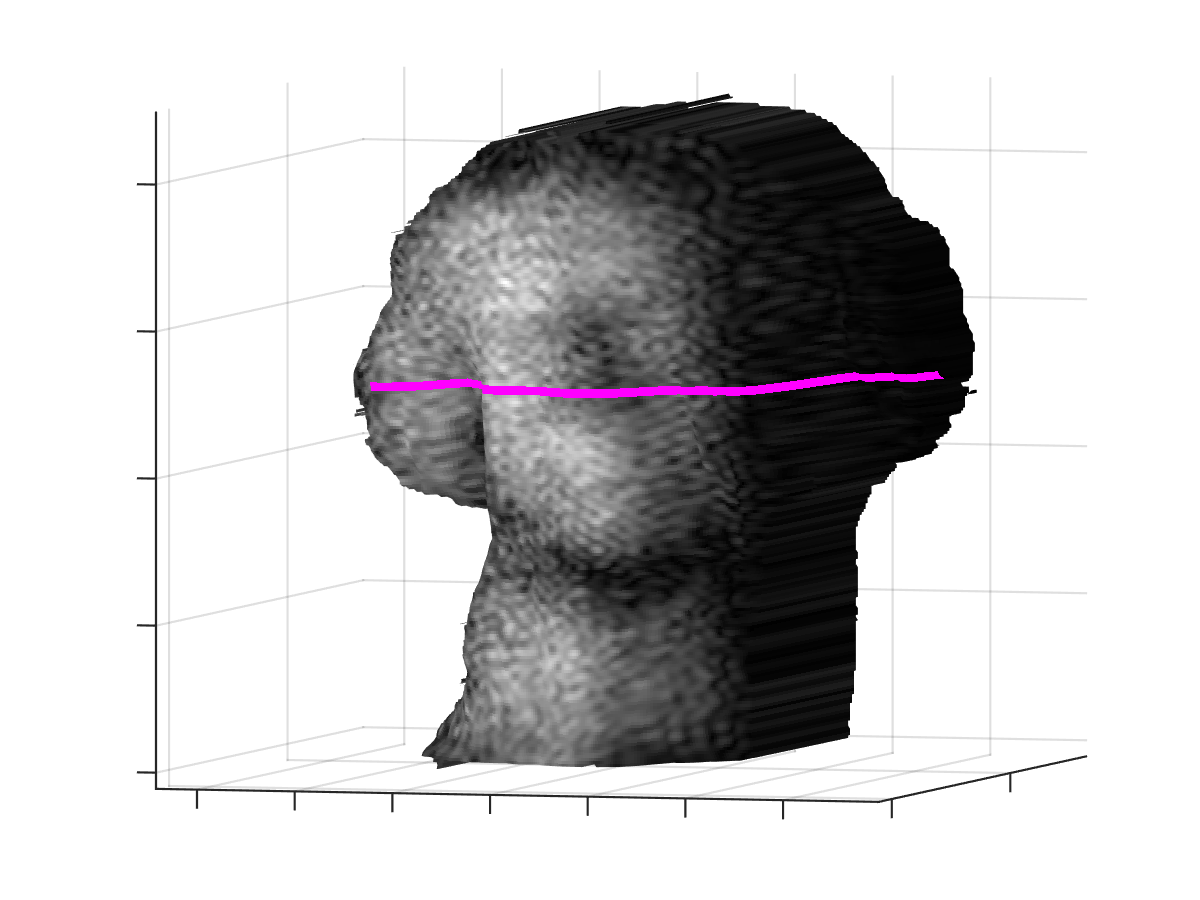}} 
  	  \centerline{(d)}
	\end{minipage}
\caption{The merging of the hologram reconstructions with the normalized depth maps. The depth map of the CGH (a), and also the depth maps of the compressive holograms with sampling rate of 50 percent (b), 25 percent (c), and 2 percent (d). The depth profile lines of the Venus statue along the frontal axis are also illustrated on the depth maps.}
\label{fig:Depth}
\end{figure}

\begin{figure}[htb!]
	\centering
	\centerline{\includegraphics[width=9.5cm]{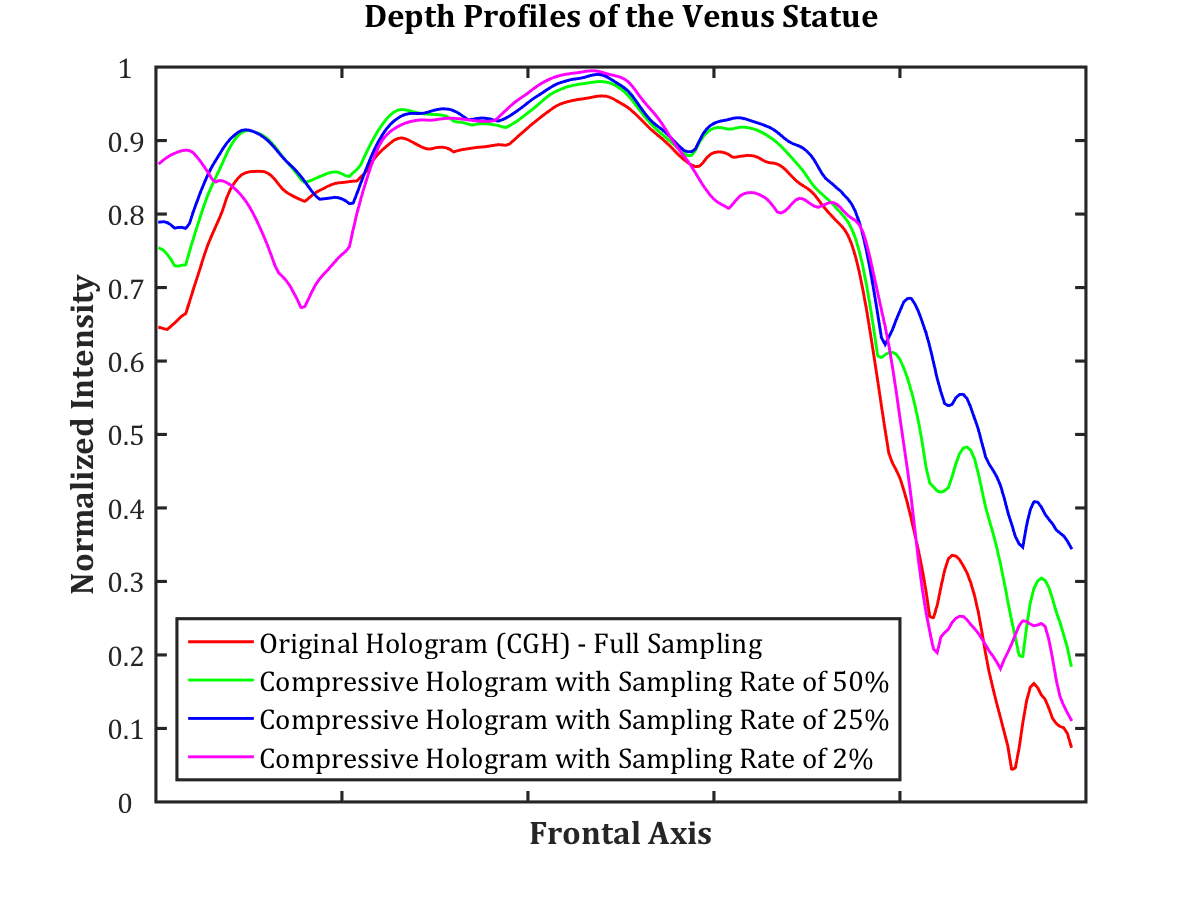}}
	\caption{The normalized depth profiles for each depth map along the frontal axis. The profile colors correspond to the colors presented on the depth maps of the Venus statue.}
\label{fig:DepthProfiles}
\end{figure}

Selection of the block size in the NCC algorithm is an important issue. The block size should be large enough for accurate matching, and it should be small enough for the less projective distortion effects. We used an empirical method to define the block size, and it was found that the best block size for our stereo image pairs was $(23 \times 23)$ in terms of estimated depth map accuracy. Once the depth maps of each hologram (compressive holograms and CGH) are acquired with the method described above, each of them is separately merged with their numerical reconstructions. The reconstructed images combined with the depth maps are illustrated in Fig. \ref{fig:Depth}. The depth profiles of the Venus statue, corresponding to the lines displayed on the depth maps in Fig. \ref{fig:Depth}, along the frontal axis are presented in Fig. \ref{fig:DepthProfiles}. The results show that the normalized depth profile of the compressive holograms with sampling rates of 2, 25, and 50 percent are very good agreement with the normalized depth profile of the CGH. These results demonstrates that it is possible to extract depth from a single compressive hologram, and that the depth extraction quality is robust to reductions in sampling rate.

\section{Conclusion}

We have presented a method that not only records a hologram faster using the compressed sensing (CS) framework but also extracts a depth map from a recorded single compressive hologram. CS can be utilized for recording holograms in a short time, since it requires a small number of measurements to acquire a scene and uses a high-speed sampling device (DMD). In addition, depth can be extracted from the compressive hologram accurately. The results demonstrate that the depth profiles of the compressive holograms are almost the same with the depth profile of the computer-generated hologram (CGH) although the hologram reconstructions are not exactly same. This shows that depth extraction does not depend on the hologram reconstruction results or sampling rates so much.

\end{document}